\documentclass[prl,twocolumn,superscriptaddress,showpacs]{revtex4}

\usepackage{amsmath}
\usepackage{epsfig}

\begin{document}

\title{Crystal Nucleation of Colloidal Suspensions under Shear}

\author{Ronald Blaak}
\affiliation{Institut f\"ur Theoretische Physik II,
  Heinrich-Heine-Universit\"at, Universit\"atsstra{\ss}e 1, D-40225
  D\"usseldorf, Germany} 

\author{Stefan Auer}
\affiliation{Department of Chemistry, Cambridge University, Lensfield
  Road, Cambridge, CB2 1EW, United Kingdom}

\author{Daan Frenkel}
\affiliation{FOM Institute for Atomic and Molecular Physics, Kruislaan
  407, 1098 SJ Amsterdam, The Netherlands}

\author{Hartmut L\"owen}
\affiliation{Institut f\"ur Theoretische Physik II,
  Heinrich-Heine-Universit\"at, Universit\"atsstra{\ss}e 1, D-40225
  D\"usseldorf, Germany} 

\begin{abstract}
We use Brownian Dynamics simulations in combination with the umbrella
sampling technique to study the effect of shear flow on homogeneous
crystal nucleation. We find that a
homogeneous shear rate leads to a significant suppression of the
crystal nucleation rate and to an increase of the size of the
critical nucleus. A simple, phenomenological extension of
classical nucleation theory accounts for these observations. The
orientation of the crystal nucleus is tilted with respect to the
shear direction.
\end{abstract}

\pacs{PACS:  82.70.Dd,  61.20.Ja, 81.10.Aj, 83.60.Rs}

\maketitle 
The formation of crystals in a supercooled melt is a
fascinating yet complex process.  It is initiated by a microscopic
nucleation event. The resulting embryonic crystal then grows to
macroscopic size.  Understanding the principles of nucleation and
growth is essential for many applications ranging from tailored
protein crystallization to
metallurgy~\cite{Kelton:1991,Galkin:2000PNAS,Shi:1995APL}.
At present, the most detailed experimental information on crystal
nucleation comes from hard sphere
colloids~\cite{Schatzel:1993PRE,Harland:1997PRE,Sinn:2001PCPS,Gasser:2001SC}.
Such suspensions are ideal to study crystal
formation, as the equilibrium and transport properties of
hard-sphere colloids are well understood~\cite{Pusey:1991}.
Moreover, recent progress in computer simulations has made it
possible to predict the absolute rate of crystal nucleation in
colloidal suspensions~\cite{Auer:2001NAT1,Auer:2002JPCM} and
thus to compare with experiment.

In the present Letter we explore the influence of shear flow on
colloidal crystal nucleation. Note that applying shear is
qualitatively different from the effect of pressure, temperature
or additives, as the latter affect the thermodynamic driving force
for crystallization or the rate of crystal growth. In contrast, a
system under shear ends up in a non-equilibrium steady state.
Several experimental studies of the effect of shear on
crystallization have been reported in the literature. Some of
these report a shear-induced ordering of the liquid which enhances
the nucleation
rate~\cite{Ackerson:1988PRL,Yan:1994PA,Haw:1998PRE,Amos:2000PRE},
while others~\cite{Palberg:1995JCP,Okubo:1999JCIS} report the
observation of shear-induced suppression of crystallization.
Both phenomena can be
qualitatively understood:  on the one hand, shear may induce
layering in the meta-stable fluid, thus facilitating crystal
nucleation. On the other hand, shear can remove matter from small
crystallites and thus works against the birth of crystals. At
present, it is not clear which mechanism is dominant, and under
what conditions. In this Letter we combine the umbrella sampling
technique from equilibrium Monte Carlo simulations with Brownian
Dynamics simulations to study this non-equilibrium problem. We
confirm that shear suppresses crystal nucleation, at least for small
shear rates, as found by Butler and
Harrowell~\cite{Butler:1995PRE}, and in addition characterize the
associated critical nucleus.

Below, we consider homogeneous crystal nucleation in a simple
model for charge-stabilized colloidal suspensions subjected to
linear shear flow. 
The charged colloidal particles
interact via a repulsive Yukawa potential~\cite{Pusey:1991}
\begin{equation}
\label{eq:pairpot}
V(r) = \epsilon \frac{e^{-\kappa r}}{\kappa r},
\end{equation}
where $\kappa$ is the inverse screening length and $r$ the mutual
distance. 
 The dimensionless
strength of the interaction  $\beta \epsilon$ has been fixed at a
value $\beta \epsilon = 1.48 \times 10^4$, where $\beta = 1/(k_B
T)$ the inverse thermal energy and we used a cut-off at a distance
$10/\kappa$. To model the time evolution of the sheared suspension, we
used Brownian Dynamics~\cite{Book:Allen-Tildesley,Chakrabarti:1994PRE}. In this
approach,  hydrodynamic interactions between the colloids are
ignored. This is justified at low volume fractions of charged
suspensions.

The Brownian-Dynamics equations of motion for a system in the
presence of a steady shear rate $\dot{\gamma}$ are of the form
\begin{equation}
\label{eq:step}
\vec{r}_i(t+\delta t) = \vec{r}_i(t) + \delta t \frac{\vec{f}_i(t)}{\xi}
+ \delta \vec{r}^G + \delta t \dot{\gamma} y_i(t) \hat{x} \;.
\end{equation}
Here $\vec{r}_i(t)=(x_i(t),y_i(t),z_i(t))$  is the position of the
$i$th colloidal particle at time $t$.  In a small time interval
$\delta t$ this particle moves under influence of the sum of the
conservative forces $\vec{f}_i(t)$ arising from the pair interaction
(\ref{eq:pairpot}) of particle $i$ with the
neighboring particles. During this motion, the solvent
exerts a friction. The friction constant $\xi$ with the solvent is
related to the diffusion constant $D$ by $\xi = k_B T /D$, while
the stochastic displacements are independently draw from a
Gaussian distribution with zero mean and variance $\langle (\delta
r_{i\alpha}^G)^2\rangle = 2 D \delta t$, where $\alpha$ stands for one
of the Cartesian components. The last term in Eq.~(\ref{eq:step}) represents
the applied shear in the $x$-direction, and imposes an explicit
linear flow field. For the simulations we used a cubic simulation
box with 3375 particles and Lees-Edwards periodic boundary
conditions~\cite{Lees:1972JPCS}. The total simulation
time was up to $10^4/(\kappa^2 D)$ for gathering statistics.
The osmotic pressure $P$ is kept at a
constant value with isotropic volume moves. In practice this means
that after a number of Brownian dynamics time steps the volume of the
simulation box is attempted to be modified and the particles locations
are scaled accordingly. The resulting difference in potential energy
is used either to accept the new volume or to reject and restore the
old volume and particle locations, following the rules as used in
normal Monte Carlo simulation of the isobaric
ensemble~\cite{Book:Allen-Tildesley}. The results for the zero shear
case show full agreement with those we obtained by equilibrium Monte
Carlo simulations.

The number of particles inside the nucleus is determined with the aid
of bond-orientational order parameters~\cite{Steinhardt:1983PRB},
which characterize the 
neighborhood of each particle. By selecting particles with a
solid-like environment that are 
in each others neighborhood, all particles that belong
to a cluster are identified.

According to the bulk phase diagram, the stable equilibrium system
would be a face-centered-cubic crystalline
phase~\cite{Auer:2002JPCM,Hamaguchi:1997PRE} for our parameters.
The system under consideration, however, is supercooled. Hence it
remains liquid, even though the solid is more stable because,
unless the nucleation rates are huge~\cite{OMalley:2003PRL}, the
simulation time required to observe spontaneous crystallization is
very much longer than the duration of a run. Due to fluctuations
the liquid will continuously form and dissolve small nuclei. Yet,
the steady state probability $P(n)$ that a critical crystal nucleus of
$n$ particles will form spontaneously is extremely small. In order to
speed up this process and obtain better statistics on the cluster size
distribution, we used the umbrella sampling
technique~\cite{Torrie:1974CPL}. 
The basic assumption underlying its usage is that the probability to
find the system with a given cluster size is a unique function of the
thermodynamic state of the system and of the shear rate. To
compute the probability to find the system in an unlikely state
(such as a critical nucleus), we bias the Brownian-dynamics
sampling in favor of the states of interest. The actual biasing
procedure is identical to the one used in (meta-stable)
equilibrium studies of crystal nucleation~\cite{Auer:2001NAT1}, and
merely works as a mathematical trick to measure the ratio of the
function $P(n)$, we want to obtain, over a known and fixed probability
$P_{bias}(n)$. All trajectories that are generated follow a normal
path and are truncated by the bias when they deviate too much from
the preferred cluster size. Rather than generating a new
configuration, the last configuration is restored from which a new
path is grown. 
Note that the trick of using umbrella sampling in a dynamical
simulation is generally applicable in equilibrium and non-equilibrium
situations, is not restricted to Brownian Dynamics, and enables one
to obtain information on rare events.

\begin{figure}[h]
\epsfig{figure=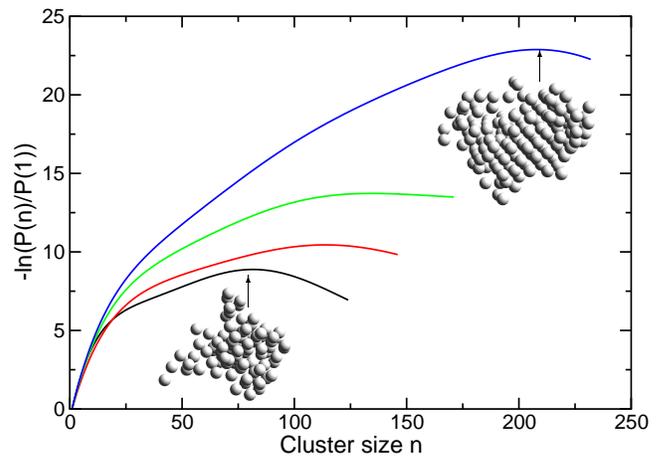,width=8.5cm,angle=0}
\caption[a]{Negative logarithm of the probability $P(n)$ of
  finding a cluster of $n$ solid-like particles normalized by $P(1)$
  for pressure $\beta P/\kappa^3 = 0.24$ and different applied shear
  rates, from bottom to top $\dot{\gamma}/(\kappa^2 D)=0$,
  $0.8 \times 10^{-3}$, $1.6 \times 10^{-3}$, and $3.2 \times
  10^{-3}$. The insets show typical snapshots of critical nuclei for
  the largest shear rate and the zero shear case.}
\label{Fig:LogP}
\vspace{-0.5cm}
\end{figure}

After correcting for the biasing function, the cluster size
distribution function is obtained. In Fig.~\ref{Fig:LogP} the
logarithm of the probability function $P(n)$ is shown for three
non-vanishing shear rates.

In the case that no shear is applied, one can relate the
probability of finding a cluster of given size to the Gibbs free
energy. It is therefore tempting to interpret the probability
functions as shown in Fig.~\ref{Fig:LogP} in terms of nucleation
barriers~\cite{Auer:2004JCP}. Strictly speaking this is not allowed,
since this idea
stems from equilibrium considerations, while in the present case
we treat a non-equilibrium system. However, application of statistical
mechanics outside equilibrium can be useful (see e.g, \cite{Ono:2002PRL}
for an effective temperature in a sheared system) and it is a
challenge to check whether and to what extent equilibrium concepts
are applicable. In our case we consider the negative logarithm of the
cluster size distribution function as an effective free energy.

Under this assumption a simple extension of classical nucleation
theory can be made, which  incorporates the shear rate. In
classical nucleation theory the Gibbs free energy
$\Delta G$ of a spherical nucleus of radius $R$ is given by
\begin{equation}
\Delta G = - \frac{4}{3} \pi R^3 \rho_S | \Delta \mu | + 4 \pi R^2 \gamma_{SL}.
\end{equation}
On the one hand there is a gain in energy proportional to the volume
of the nucleus due to the difference in chemical potential $\Delta \mu$
between the solid with density $\rho_S$ and the liquid phase. On the
other hand we have a loss in energy, since an interface between the
solid nucleus and surrounding liquid needs to be formed, described by
$\gamma_{SL}$ the interfacial free energy.

It is reasonable to expect that for moderate shear rates the
chemical potential difference $\Delta \mu$ and interfacial free
energy $\gamma_{SL}$ will not be affected much. This would justify
an expansion in powers of the shear rate for both these quantities
about their equilibrium values
\begin{equation}
\label{eq:mugamma}
\begin{split}
\Delta \mu & = \Delta \mu^{(eq)} \left(1 + c_0 \dot{\gamma}^2 + {\cal
  O} (\dot{\gamma}^4) \right) \\
\gamma_{SL} & = \gamma_{SL}^{(eq)} \left(1 + \kappa_0 \dot{\gamma}^2 +
  {\cal O} (\dot{\gamma}^4) \right),
\end{split}
\end{equation}
where due to the invariance of the shear direction only even
powers in the shear rate $\dot{\gamma}$ need to be considered.

\begin{figure}[ht]
\vspace{-0.2cm}
\epsfig{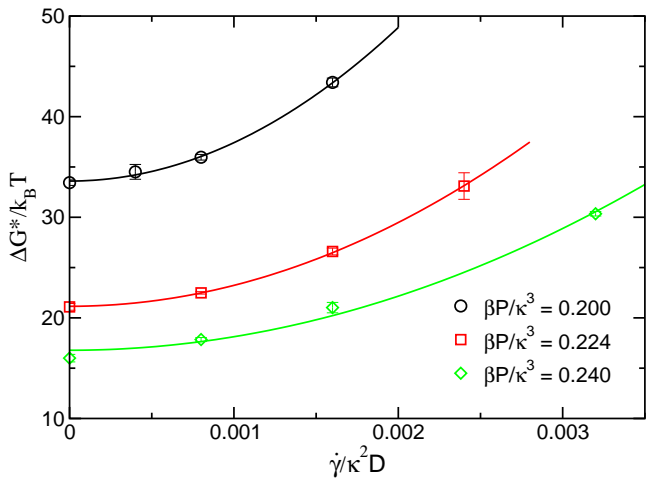}
\caption[a]{The height of the nucleation barrier $\beta \Delta G^*$ as
  function of the dimensionless shear rate $\dot{\gamma}/\kappa^2 D$
  for different pressures $P$. The solid lines are parabolic fits
  through the data.}
\label{Fig:dG}
\vspace{-0.3cm}
\end{figure}

If we combine these expansions with the expression from classical
theory one can easily derive expressions for $\Delta G^*$, the
height of the nucleation barrier and $N^*$, the size of the
critical nucleus,  both of which depend quadratically on the shear
rate. In Fig.~\ref{Fig:dG} we show the results from our
simulations where we extracted the height of the nucleation
barrier for various pressures and shear rates. The dependence on
the shear rate is confirmed by the parabolic fits. However, we
caution the reader that this observation should not be considered
as evidence that the shear rate can really be considered as a
thermodynamic variable. In fact, in a recent study of the effect
of shear on the location of the solid-liquid coexistence in a
Lennard-Jones system, Butler and Harrowell found that no purely
thermodynamic description of the effect of shear was
possible~\cite{Butler:2003JCP}. Shear directly affects the
transport of particles from the solid to the liquid phase, and
this effect is not thermodynamic. The expansion in
Eq.~(\ref{eq:mugamma}) is simply a way to represent the effect of
shear as if it were purely thermodynamic. With this caveat in
mind, we continue the remainder of the discussion in the language
of classical nucleation theory.
We find that $N^*$, the number of particles inside the critical
cluster, also depends quadratically on the applied shear rate.
Using the classical nucleation theory expressions $N^*=
(32\pi\gamma^3)/(3 \rho_S^2|\Delta\mu|^3)$ and $\Delta G^* =
N^*|\Delta\mu|/2$, we can obtain the values of the second order
coefficients in Eq.~(\ref{eq:mugamma}) from a fit of the simulation
data.  The results are summarized in Table~\ref{Tab:data}. We find a
negative $c_0$ implying a destabilization of the solid upon shear and a
relatively small correction of the interfacial free energy. Both
effects do not strongly depend on the pressure. Note, however, that
the fits for $\Delta\mu$ and $\gamma$ do not yield a good prediction
for the {\em shape} of the nucleation barrier. 
The shape shows deviations from the one expected by classical
nucleation theory, which is due to finite size effects of the cluster.
\begin{table}[ht]
\vspace{-0.1cm}
\begin{tabular}{c|c|c|c|c}
$\beta P/\kappa^3$ & $\beta \Delta G^{(eq)}$ &
$N^{(eq)}$ & $c_0 D^2 \kappa^4$ & $\kappa_0 D^2\kappa^4$ \\
\hline
0.200 & 34 & 209 & -4.8 $\times$ 10$^4$ & 6. $\times$ 10$^3$ \\
0.224 & 21 & 133 & -4.1 $\times$ 10$^4$ & 5. $\times$ 10$^3$ \\
0.240 & 17 &  97 & -3.4 $\times$ 10$^4$ & 4. $\times$ 10$^3$
\end{tabular}
\caption{Numerical data for different pressures $\beta P/\kappa^3$ on
  the equilibrium barrier height $\Delta G^{(eq)}$, critical nucleus
  size $N^{(eq)}$, and second order corrections to the free energy
  difference and interfacial free energy as obtained from the fitted
  simulation data.}
\label{Tab:data}
\vspace{-0.4cm}
\end{table}

A bond order analysis shows that
the structure of the nucleus is predominantly body-centered-cubic.
 Since small nuclei are in general neither spherical nor
compact we have chosen to characterize their shape by the three
principal moments of inertia. For a truly spherical nucleus these
values would be identical, but since the shape of nucleus is
fluctuating these moments are different. For relatively small
clusters of 100 particles the ratio of the principle moments is
roughly 6:10:12.  As the nuclei grow larger, the differences
between these moments of inertia get somewhat less.  Surprisingly,
the imposition of shear  does not influence these ratios. This
leads to the conclusion that although the size of the critical
nucleus increases with shear, the overall shape is hardly
influenced. This is different from the radial distribution functions
we measured in the liquid under shear. They become increasingly
asymmetric for higher shear rates~\cite{Blaak:2004JPCM}.

Knowledge of the eigenvectors of the inertia tensor allows us to
determine its orientation. We find that the average orientation of
the nucleus is weakly coupled to the direction of the applied
shear. In particular, we find that the axis with the largest
principal moment of inertia is, preferably in the gradient direction,
in qualitative difference to a typical nearest neighbor particle
cluster in a sheared fluid that prefers to be in the shear
direction. The axis of the smallest principal moment of the nucleus
tends to align with the vorticity direction. This alignment becomes more
pronounced with increasing nucleus size and with increasing shear
rate.

\begin{figure}[ht]
\vspace{-0.2cm}
\epsfig{figure=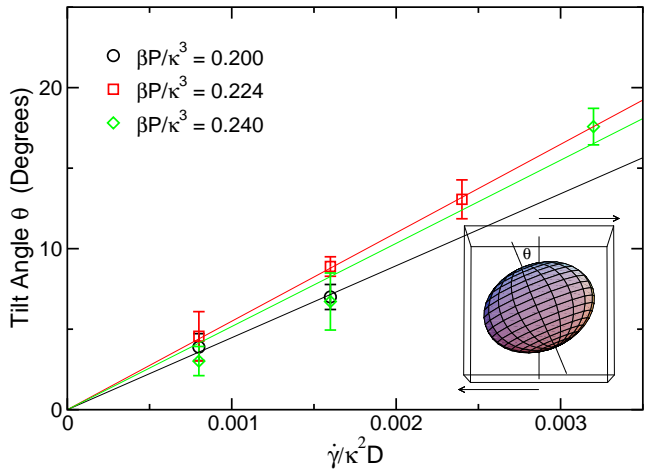,width=8.5cm,angle=0}
\caption[a]{The tilt angle $\theta$ of the principal moment of inertia
  with respect to the $y$-axis.
  The inset shows a schematic
  representation of the preferred orientation of the nucleus with
  respect to the shear direction indicated by the arrows.
}
\label{Fig:orient}
\vspace{-0.3cm}
\end{figure}

In Fig.~\ref{Fig:orient} we show the orientation of the nucleus
with respect to the shear direction. The tilt angle increases linearly
with the applied shear rate $\dot{\gamma}$ and only depends weakly on
the osmotic pressure. In order to improve the statistical accuracy we
have averaged over all cluster sizes between  $N=100$  and the
critical nucleus size, $N^*$. The inset of Fig.~\ref{Fig:orient}
shows a schematic drawing of the preferred orientation of a
nucleus. Note that the largest dimension of the nucleus (smallest
principle moment) is preferably along the vorticity direction,
i.e.~perpendicular to the plane of drawing. Interestingly, a similar
tilt occurs when vesicles with a flexible shape are exposed to a
linear shear flow~\cite{Abkarian:2002PRL}.

In conclusion, we applied the combination of umbrella sampling and
Brownian Dynamics simulation to the non-equilibrium problem of
nucleation under shear, and found that shear suppresses nucleation and
leads to a larger critical nucleus.
These results can be described (but not yet
understood) using a naive extension of classical nucleation
theory. Most importantly, the present numerical predictions can be
tested experimentally, by studying the rate of homogeneous crystal
nucleation in a homogeneously sheared colloidal suspension. If
nucleation were to be studied in Poiseuille flow as realized in a
capillary viscometer~\cite{Palberg:1996JP1F}, rather than in
homogeneous Couette flow, we should expect crystal nuclei to
appear preferentially in the middle of the flow channel.

We stress that the present findings apply to the case where the
fluid is only weakly sheared, i.e.~when shear-induced ordering in
the liquid phase is, presumably, unimportant. We also note that
the present results indicate that, during sedimentation of crystal
nuclei in an otherwise stagnant solution, local shear should
decrease the rate of growth of the crystallites. There may even be
conditions where the competition between mass gain due to crystal
growth and mass loss due to shearing, leads to the selection of
one particular crystallite radius. This
phenomenon should also be experimentally observable.

In this work, we ignored hydrodynamic interactions because
otherwise the computational cost would have been prohibitive. This
assumption, while reasonable for dilute suspensions of charged
colloids, is certainly not correct in general. Finally, our method
can readily be applied to other dynamical simulation methods for rare
events and meta-stable systems, such as crystal nucleation in oscillatory
shear~\cite{Xue:1989PRA} and heterogeneous nucleation near a
system wall in a sheared suspension.

\begin{acknowledgments}
We like to thank T. Palberg, A. Van Blaaderen, G. Szamel, and S.
Egelhaaf for helpful discussions. This work has been supported by
DFG within subproject D1 of the SFB-TR6 program. The work of FOM
Institute is financially supported by the
``Nederlandse organisatie voor Wetenschappelijk Onderzoek'' (NWO).
\end{acknowledgments}

\end{document}